# Perfect plasticity of metals under simple shear as the result of percolation transition on grain boundaries

Y. Beygelzimer[a*], N. Lavrinenko[b]

[a]Donetsk Institute for Physics and Engineering named after O.O. Galkin of the National Academy of Sciences of Ukraine, 72 R. Luxemburg St., 83114 Donetsk, UKRAINE

*Corresponding author. Email: yanbeygel@gmail.com

[b]Scientific and Technologic Center "Reactivelectron" of the National Academy of Sciences of Ukraine, 17a, Bakinskih komisariv St., 83049, Donetsk, UKRAINE

A mechanism of perfect plasticity under low homological temperatures has been suggested. According to this mechanism, the phenomenon under study is of critical nature. It connects with percolation transition in the net of grain boundaries and with nonlocal interaction of fragments uniquely under simple shear mode. The mechanism is justified by general reasoning, mainly of geometrical character, and also by employing computational modeling and well-known experimental results.

## 1. Introduction

At the present time it is reliably established that under sufficiently large deformation of metals under simple shear mode and low homological temperatures the stage of perfect plasticity comes. That is, the shear stress reaches some fixed level of saturation and does not increase any more during simple shear loading.

P. Bridgeman was one of the first to reveal abnormally low hardening under large strain in his classical high pressure torsion experiments [1]. Later, multiple investigations (see, for example, [2, 3]) confirmed that torsion test diagram exhibit saturation. Finally, most convincingly perfect plasticity was detected in recent investigations of high pressure torsion [4-7]. These studies did not only show that torque was constant under given enough shear strain, but also detected that the microstructure of specimen remained the same.

Perfect plasticity under large deformations indicates that qualitatively new state of metals is observed. Hence, similarly to superplasticity, superfluidity, and superconductivity, this phenomenon is of fundamental interest. It was the flow of solid bodies "like liquids" that impressed A. Treska the most in his historically pioneer investigations of plastic deformations [8]. However, situation developed in such a way that "perfect plasticity" for a long time remained just a mathematical model for the scientists, while attention was attracted to elastoplastic transition.

In connection with perfect plasticity phenomenon two questions arise: what causes it and whether it can be observed not only during deformation according to simple shear scheme, but also under other schemes of loading?

There are no definitive answers on the posed questions hitherto. So, basing on the results of experiments conducted, P. Bridgeman supposed that abnormally low hardening occurs only under simple shear deformation scheme. In order to justify this idea he suggested an idealized nuclear model for metals deformation. The model illustrated fundamental difference between simple shear scheme and stretching one [1]. The author [9] also considered that loading scheme affects structure and characteristics of metals under large deformations.

On the other hand, according to [4], experimental results persuasively show that, no matter what the deformation mode is, evolution of metal structure mainly follows universal regularities. These regularities were established in [10] and [11]. That is the reason why the authors suggest that perfect plasticity phenomenon should be common for different loading modes though it was experimentally proved only during high pressure torsion. The problem was to achieve large deformations under other loading schemes. This point of view is shared by authors of [5] – [7].

In this article we suggest a mechanism of perfect plasticity under low homological temperatures. According to this mechanism, the phenomenon under study is of critical nature. We consider that it is connected with percolation transition in the net of grain boundaries and with nonlocal interaction of fragments uniquely under simple shear. Our point of view is substantiated by general reasoning, mainly of geometrical character, and also by employing computational modeling and well-known experimental results.

The article develops concepts [12-15], which are based on the hypothesis of the first author that during large deformations according to simple shear scheme turbulent flows in metals occur.

## 2. Special property of simple shear and its relation with perfect plasticity

Simple shear is given by:

$$\begin{aligned} x^1 &= X^1 + \gamma \cdot X^2 \\ x^2 &= X^2 \\ x^3 &= X^3 \end{aligned} \quad (1)$$

where $X^i$ and $x^i$ are correspondingly initial and end coordinates of material point $(i = 1, 2, 3)$; $\gamma$ is shear strain; $X^2 = 0$ gives a plane of shear deformation and direction of the shear coincides with the $X^1$ axis.

We will show if the pressure remains constant under simple shear of the material, this material is necessarily perfectly plastic.

Let us perform a thought experiment. Imagine a round disk of deformed specimen that is clamped between two rigid anvils, the upper of which is being twisted and the lower one is fixed (Figure 1). There is no slipping between the anvils and the specimen. In cylindrical coordinates system (r, φ, z) the velocity field with the following components is realized: $V_r = V_z = 0$, $V_\varphi = rz\omega$, where $\omega = \dot\varphi = \text{const}$, and a dot over φ means time derivative.

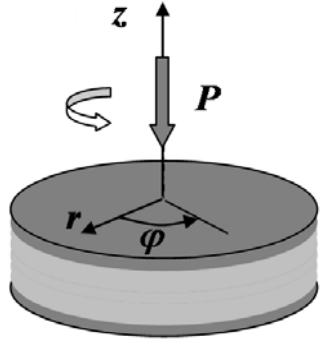

Figure 1. Deformation scheme for a specimen material layer clamped between two rigid anvils.

Under such conditions at any moment of time in a small neighborhood of any point of the disk simple shear deformation occurs. The stressed state is described with tangent stress $\tau$ and pressure $p$.

From dimensional considerations, it appears that $p(r,\gamma) = \sigma(\gamma) \cdot f(r,\gamma)$, where $\sigma(\gamma)$ the flow stress of the deformed material is, $f(r,\gamma)$ is a dimensionless function of the shear strain $\gamma$ and distance between the point and the $z$ axis. Relation between the anvils pressing force $P$ and $p(r,\gamma)$ is given by:

$$P(\gamma) = \int_S p(r,\gamma) dS = 2\pi\sigma(\gamma) \int_0^R f(r,\gamma) r dr, \qquad (2)$$

where $R$ is the disc radius.

Let us consider the system for two shear strain values $\gamma_1$ and $\gamma_2$. As the system is symmetric with respect to the rotation about axis $z$ (see Figure 1), it holds that $f(r,\gamma_1) = f(r,\gamma_2)$. Then from relation (2) it follows that under constant anvils pressing force $P(\gamma_1) = P(\gamma_2)$ the yield stress of the material and pressure at any point of the disc are also constant, that is $\sigma(\gamma_1) = \sigma(\gamma_2)$ and $p(r,\gamma_1) = p(r,\gamma_2)$.

Thereby, we can conclude that simple shear under constant pressure should pass without hardening of the material. It is a necessary condition. It follows from the condition that if simple shear is possible under constant pressure then such material has to be perfectly plastic. And when the material is hardening under deformation it becomes necessary to continuously increase the pressure in order to realize simple shear scheme. Simple shear mode under constant pressure is impossible for a hardening material. This result was obtained for the first time in [13, 14], where the agreement of the result with experiments was also shown.

The principal property of the simple shear on which we based our reasoning is that under transition (1) different material points move in a parallel way relative to each other. It is this property that leads to invariance of the geometry of the system in consideration with respect to the rotation.

If we consider a thin layer of material along the motion direction in a neighborhood of any point of the specimen, cross-section size of the layer does not change under deformation. Owing to this fact, a stationary structure may be formed in the material and such a structure results in perfect plasticity. Let us show that if the deformation scheme does not possess this specified property then stationary structure cannot emerge.

Consider a pure shear – flat lengthening deformation without change in volume. This transformation is given by:

$$\begin{aligned} x^1 &= kX^1 \\ x^2 &= k^{-1}X^2 \\ x^3 &= X^3 \end{aligned} \quad , \tag{3}$$

where $k$ is the elongation coefficient along the $X^1$ axis.

Suppose that for some value $k^*$ the cross-section size of the specimen along the $X^2$ axis is equal to $a$ and the average cross-section size of microstructure fragments is $d$. Assume that starting from this moment further deformation does not lead to microstructure changes (steady-state process).

Extending deformation, the elongation coefficient can be increased to some value $k^{**} > k^* \frac{a}{d}$. Then from (3) it follows that after deformation the cross-section size of the specimen should become less than the average size of the microstructure fragments is. But it is a contradiction. Therefore, stationary microstructure under flat lengthening deformation mode is impossible.

Certainly, because of a number of reasons (such as dynamical recrystallization, migration of grains boundaries, boundaries sliding, etc.), at some stages of deformation process the cross-section size of fragments may decrease more slowly than the cross-section size of the specimen. However, our considerations imply that under lengthening deformation scheme the process of fragmentation has to recrudesce until the specimen breaks or stretches to one-dimensional chain of indivisible fragments, atoms in the limit. Consequently, under lengthening deformation of the specimen and deformation of the microstructure are similar, which is reflected by the Polanyi-Taylor principle [16].

So we showed that under simple shear mode there are all necessary prerequisites for perfect plasticity and stationary microstructure emergence. How do they emerge and sustain during the deformation? Do there just universal regularities take place [10, 11, 17] or can there appear some new mechanism? In the next section we suggest a geometrical model for plastic deformation of metals that may help to answer these questions.

3.  **Plastic deformation of metals as isometric transition with singularities**

Real metallic specimens are large constructions with a huge number of atoms. Deformation of a specimen is related to a change of atoms positions in space and can be described by transformation of their coordinates:

$$\mathbf{m} = \mathbf{G}(\mathbf{M}) \ , \qquad (4)$$

where $\mathbf{M}$ and $\mathbf{m}$ are correspondingly the vectors of initial and end coordinates of atoms.

Because of high dimensionality of the problem, it is practically impossible to determine the transformation in (4) and relate it to the loading that was applied to the specimen. That is why deformation process is usually considered at several multi-scaled levels.

At the macro level metals are modeled as a solid continuous medium, deformation of which results in changes in lengths of material fibres and in angles between them. This can be given by affine transformation:

$$d\mathbf{x} = \mathbf{F}(\mathbf{X})d\mathbf{X} \qquad (5)$$

where $\mathbf{X}$ and $\mathbf{x}$ are coordinates of a material point before and after deformation; $\mathbf{F}(\mathbf{X}) = \dfrac{d\mathbf{x}}{d\mathbf{X}}$ is the deformation gradient tensor.

Transformation (5) in fact represents transformation (4) at the macro level without taking in regard micro-scale effects.

At the micro level metals possess crystalline lattice that can bear only reversible elastic deformation with the order of magnitude not more than $10^{-3}$. Therefore, at the micro level transformation (4) can be considered to be isometric, that is, without changes in lengths of segments. Such transformations include translation, rotation, and symmetric reflection.

According to the theorem stated in [18], continuous transformation that is near-isometric in the small neighborhood of any point is isometric in the whole region. That is why transformation (4) can change lengths of segments at the large-scale level only if it is isometric transformation with singularities (piecewise isometric transformation) [19]. The latter represent surface of isometric discontinuity, owing to which large values of derivatives in transformation (5) are achieved.

It can be seen that large plastic deformations in metals are realized only when discontinuities in isometric transformations happen. Dislocations, grain boundaries dislocations, disclinations, and twins become bearers of such discontinuities. Surfaces that sweep the discontinuities under their motion form a set of singularities of an isometric transformation (4).

The fact, that transformation (4) belongs to the class of isometric transformations with singularities, gives a key to answer the questions we posed in the previous section. Let us demonstrate it.

We will employ a simple model (Figure 2) to see how singularities of isometric transformations emerge.

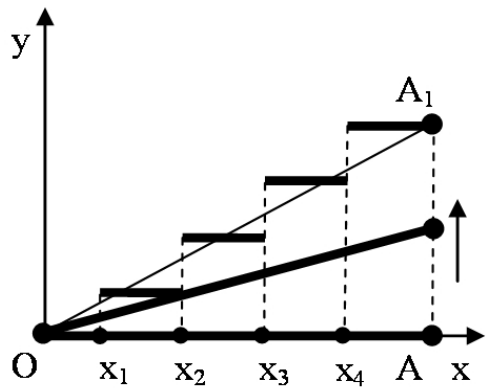

Figure 2. "Lengthening" of the segment OA by the way of isometric transformation with singularities

If the point A shifts upright, OA segment becomes longer and deviation from isometry occurs. When this deviation reaches its critical value (point A reaches position A1), ruptures at points $x_i$ $(i=1,...,4)$ emerge and restore isometry at the small-scale level, however allowing the whole segment to "lengthen". Further motion of point A upright will lead to subsequent reiteration of the process – deviation from isometry superseded by local restoration of isometry due to emergence of new discontinuities in the isometric transformation.

From our reasoning it appears that transformation (4) can be given in the following way:

$$\mathbf{G} = \prod_i \Delta \mathbf{G}_i \qquad (6)$$

where $i$ is the deformation step,

$$\Delta \mathbf{G}_i = \Delta \mathbf{P}_i \Delta \mathbf{E}_i, \tag{7}$$

$\Delta \mathbf{E}_i$ are affine transformations for a small elastic deformation of the crystalline lattice; $\Delta \mathbf{P}_i$ are piecewise isometric transformations that bring parameters of the lattice back to initial values and allow the representative volume of material to accomplish large deformation.

Transformations $\Delta \mathbf{P}_i$ result in periodical relaxation of elastic stress at the micro level because a set of discontinuities of isometry emerges, we will designate this set $D_i$. It is these transformations that describe the structural evolution based on a number of governing principles [10, 11, 17].

So the issue of stationary microstructure raised in the previous section now can be connected with the search for a stationary piecewise isometric transformation $\Delta \mathbf{P}$ that is a self-mapping of $D$. Its repeated application will not enhance the surface of discontinuity of isometry. Rotation of a circle on a plane is such transformation (Figure 3).

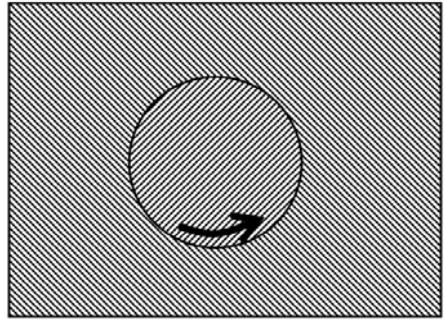

Figure 3. Stationary isometric transformation with singularity on the circumference.

In the following section we will show that simple shear in metals can to certain conditions be realized in a similar way.

4. **Percolation model of a stationary microstructure under simple shear scheme**

Let us consider schematically a grained refinement process. According to [10, 11, 17], at the initial stage of deformation weakly misoriented cells the size of about dozens of nanometers emerge, forming a fine-meshed net consisted of small-angle grain boundaries. Starting from some moment, areas with high-angle misorientations appear in the net and their number increases as deformation continues.

It is widely known that sliding along high-angle boundaries caused by deformation is possible (even at cryogenic temperatures) [20, 21]. In this case such boundaries constitute the set of isometry discontinuities $D$. Sliding occurs by movement of grain-boundary dislocations.

For the purpose of further discussion, boundaries from the set $D$ would be convenient to represent as a series of separatory rolls, rotation of which leads to relative shift of adjacent areas. We will consider only plane problems, so the boundaries can be given as depicted in Figure 4.

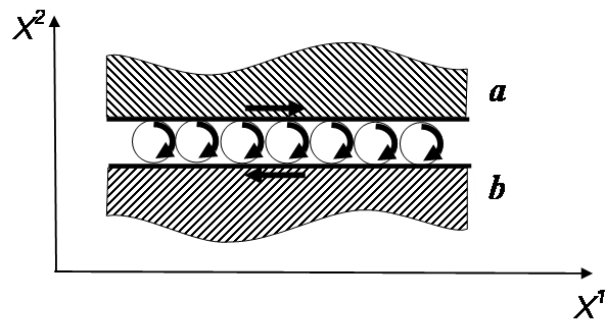

Figure 4. Model for the discontinuity boundary in a form of chain of rolls, rotation of which leads to relative shift of areas a and b.

Rolls can rotate clockwise as well as counterclockwise. Easy to see that in the first case they correspond to positive values of $\frac{\partial x^1}{\partial X^2}$ components of the deformation gradient and negative values for $\frac{\partial x^2}{\partial X^1}$ components, in the other case – on the contrary.

The above introduced scheme of the fragmentation process makes possible to describe it in percolation theory terms (bonds problem) [22] as a consecutive transformation of some lattice, resulting in that more and more of its parts becomes elements of the set $D$.

Let us introduce $\Theta$ - relative part of the lattice elements, belonging to the set $D$. According to the percolation theory, while $\Theta \ll 1$, such elements are individual random inclusions in the lattice. With the growth of $\Theta$ they start to form interconnected groups – clusters. When $\Theta$ reaches some critical value $\Theta_c$ qualitative change occurs – emerges the so called percolation cluster (PC) penetrating through the whole lattice. $\Theta_c$ is named a percolation threshold and depends on the type of the lattice. For example, for a hexagonal lattice $\Theta_c = 0.65$.

As percolation cluster penetrates through the entire lattice, it determines isometry discontinuities emerging in the whole representative volume. Starting from that moment, transformation $\Delta \mathbf{P}_i$ in representation (7) can proceed mainly through shear along the boundaries of PC. The latter has a fractal structure with loops of different scales [22]. Average crosscut size of the loops $L$ equals the correlation radius that is much larger than the cell's size $l$. For this reason at the scale of $L$ the loops seem to be smooth. Value $L$ near the percolation threshold can be written as [22]:

$$L = l \left| \Theta - \Theta_c \right|^{-\nu}, \tag{8}$$

where $\nu$ is the index for correlation radius. For two-dimensional problem $\nu = 1.33$.

In Figure 5 one step of simple shear under transformation (7) with percolation cluster is described. The latter is schematically shown as hexagonal lattice with cells sized $L$. According to (1), under simple shear it is necessary to keep $\dfrac{\partial x^1}{\partial X^2}$ positive. It means that in the model from Figure 4 the rolls rotate clockwise. Therefore, the areas bounded by cells of the percolation cluster will rotate counterclockwise, which is shown in Figure 5. From Figure 5 it can be seen that in those areas isometry sustains to the first approximation and deformation gradient of simple shear is achieved mainly by the way of shifts along the percolation cluster.

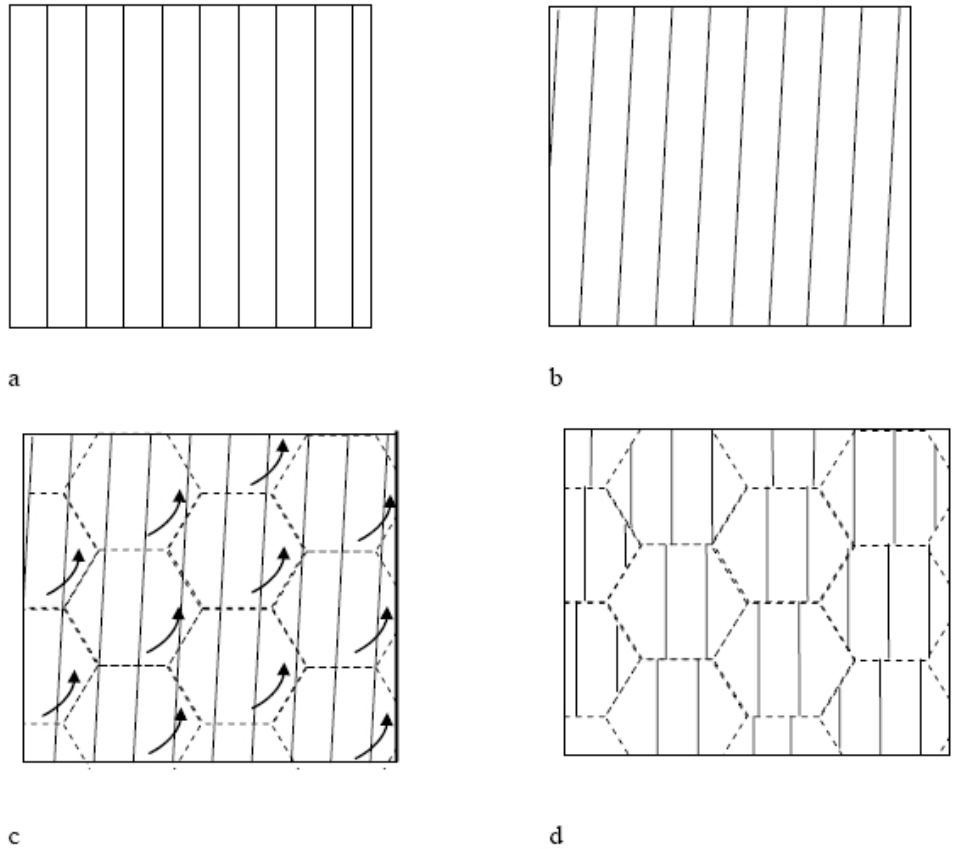

Figure 5. Scheme of the simple shear under transformation (7) with the presence of a percolation cluster: a – initial state; b – $\Delta E$ transformation result (elastic deformation); c – percolation cluster is outlined by dashed line; d - $\Delta P$ transformation result based on shift along the boundaries of the percolation cluster.

We considered only one step of a simple shear. All further steps are similar, so steady state of the proposed transformation is sustained (in statistical sense).

It should be noted that each step is related to the effect of its own percolation cluster. The matter is that the motion of grain boundary dislocations along the border leads to the change of the grained boundary structure that may temporary impede the next shift (stick-slip effect [23]). So in the result of one step of transformation (6) the percolation cluster may break in several points, so that at the next step another percolation cluster would be needed. It means that in order to realize the suggested mechanism there should be some reserve of lattice elements belonging to the set $D$. In other words, shear percolation that started at $\Theta = \Theta_c$ would become stationary at $\Theta = \Theta_{c1} > \Theta_c$.

Let us specify main properties of the suggested mechanism of simple shear.

(a) The suggested mechanism is not of local, but of cooperative character with correlation radius $L$. At the scales of sizes less than $L$ percolation cluster is of fractal nature, for the scales higher than $L$ it is homogeneous (crossover effect, [22]). According to relation (8), near the percolation threshold the correlation radius sharply increases. For $L \geq H$ (where $H$ is the thickness of the shear layer, e.g. distance between anvils in high pressure torsion) percolation cluster may break in the direction of the shear. It means that for a regular percolation, that is for perfect plasticity, the value of $L$ has to decrease. According to (8), it requires an increase in $\Theta$. It appears that the percolation threshold $\Theta_c(H)$ in the layer of thickness $H$ exceeds the threshold $\Theta_c$ in the system of infinite scale. From (8) a simple relation between percolation thresholds for layers of thicknesses $H_1$ and $H_2$ can be derived:

$$\frac{\Theta(H_1) - \Theta_c}{\Theta(H_2) - \Theta_c} = \left(\frac{H_2}{H_1}\right)^\nu \qquad (9)$$

(b) It can be seen in Figure 5 that sequential application of transformation (7) for different percolation clusters results in stirring of the material. In [14] the upper estimate for $R$ mean-square shift of fragments under the mechanism is assessed:

$$R = \sqrt{\frac{Ll}{2}\gamma} \qquad (10)$$

By substitution of (8) the following can be written:

$$R = l\sqrt{\frac{\gamma}{2}}|\Theta - \Theta_c|^{-\frac{\nu}{2}} \qquad (11)$$

So, fast mass transfer is the consequence of the suggested mechanism of simple shear.

(c) Under transformation (7) orientation of the fragments remains persistent (Figure 5). For this reason in the process of stirring strongly disoriented and weakly disoriented fragments contact themselves. In the second case they "stick" and coarsen. The coarsened fragments then break again. All the above supports the idea that under the suggested mechanism orientation of fragments is dynamically persistent, as well as their average size and distribution of sizes. Values of theses parameters correspond to the ones at the beginning of percolation.

(d) The piecewise isometric transformation that was discussed in this section is basing on rotations of volume areas, bounded by percolation cluster cells. As the latter have fractal structure at the scales less than $L$, so it is multi-scale rotations that are in some way similar to the ones that emerge under turbulent liquid flows [12, 13]. However the reason for such rotations in solids differs from the reason for turbulence in liquids.

In the next section we will state and substantiate a hypothesis that the driving force for rotations is couple stresses emerging in the representative volume of material under simple shear.

## 5.    Couple stress as the driver for rotation

Classical theory of continuous media is based on a hypothesis of symmetrical stress tensor. In the vast majority of practically important cases this hypothesis is consistent with experimental data. Significant deviations from experimental results arise when stress gradient is large. In particular, it is the case of polycrystals [24] or grain media. Because of essential inhomogeneity there arise sharp stress drops which result in effects that symmetry theory cannot describe. To study such materials nonlocal mechanics is employed [25].

From our point of view, such effects take place in metals during simple shear mode near the described above percolation transition. They are associated with the violation of the shear stress reciprocity law at the $L$ scale. Vacancies emerging in the border areas may become the reason for this law violation during simple shear. Indeed, increasing number of vacancies causes an increase in volume of material in the boundary area. Additional work needed to shift against the pressure leads to an increase in tangential stress [26].

Schematically, let the percolation cluster cell be a square of size $L$ (Figure6a). Consider the forces that act on the square. Let the pressure along the $z$ axis exceed the pressure along the $x$ axis. Pursuant to the above reasoning, such a situation will lead to an inequality $\tau_{zx} > \tau_{xz}$, that is, to the violence of shear stress reciprocity law at the $L$ scale.

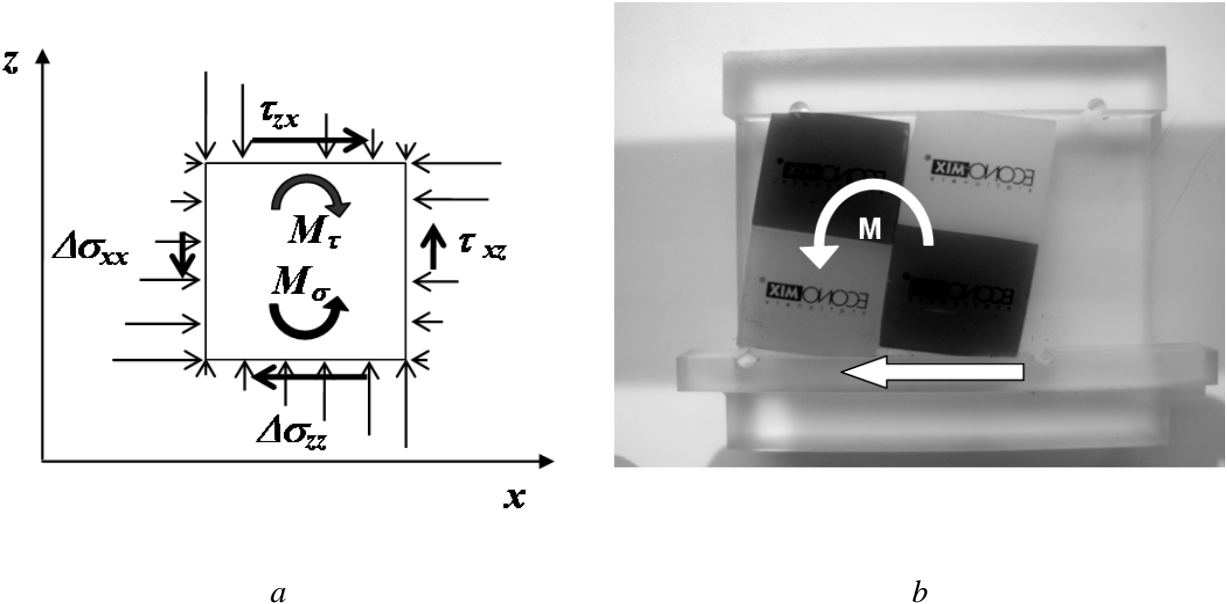

*a*  *b*

Figure 6. Model describes torques emergence under simple shear scheme: a – the scheme of forces and torques applied to the PC cell; b – model with four rubbers between perspex plates demonstrates how elastic torques emerge when the lower plate moves and generates frictional stress differential on the horizontal and vertical sides of rubbers.

From Figure 6a it comes that the force moment affecting the square cell, which is caused by tangential stress, is the following: $M_\tau = L^2(\tau_{zx} - \tau_{xz})$ (assume that the thickness of the cell in the direction perpendicular to the xz plane is unit). Its small rotation, caused by the moment, results in respondent elastic reaction of the surrounding material. This reaction can be described by an inhomogeneous field of normal stresses $\Delta\sigma_{xx}$ and $\Delta\sigma_{zz}$ that creates a compensating moment $M_\sigma$ affecting the cell. The case is described in Figure 6a. A simple physical model in Figure 6b illustrates the effect.

While the moments are equal the cell is equilibrium. Under a certain threshold value $\tau_{zx}$ conditions favoring stick-slip effect emerge (mentioned above). This leads to an imbalance between $M_\tau$ and $M_\sigma$, causing a stepwise rotation of the cell under the force moments difference $M_\sigma - M_\tau$. Then comes relaxation $M_\sigma$ and equilibrium restores.

The aforesaid is explained by a mechanical model represented in Figure 7.

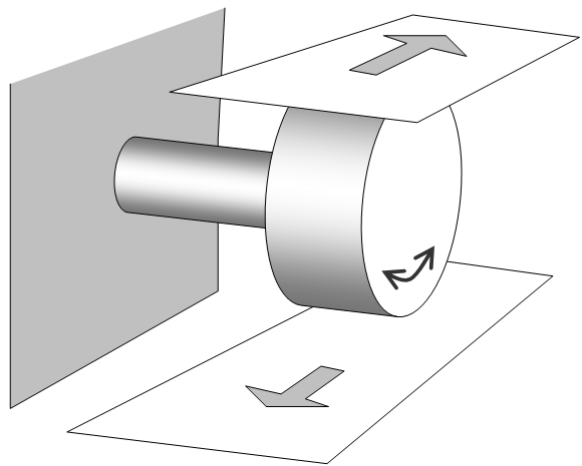

Figure 7. Mechanical model that explains the mechanism of deformation under simple shear

A disc clamped between two parallel plates is fixed to the wall by means of a elastic cylinder. Small shift of the plates in opposite directions generates a moment of frictional force applied to the disc. It causes a elastic response of the twisted disc. It is known that due to sticking and slipping in the contact area between the disc and the plate shifting the plates causes the so-called frictional self-oscillation of the disc with periodical relaxation of the elastic moment of the disc in the model [23].

Let us show that the above described mechanism of stress relaxation is typical for simple shear scheme only. For this purpose consider loading of a mechanical construction from the idealized model in Figure 3.

The construction consists of a flat plate of height $H$ and width $B$, containing a cylindrical inclusion of diameter $D$. Under deformation of the construction slipping between the inclusion and the plate is possible, but no gaps are allowed. At the border between the elements holds the following:

$$\tau \leq \tau_0 + \mu p, \qquad (12)$$

where $\tau$ and $p$ are correspondingly tangential stress and pressure along the border; $\tau_0$ and $\mu$ are the cohesion sliding resistance and friction coefficient between the plate and the inclusion. Along the borders, where equality in (9) holds, slipping between the inclusion and the plate occurs.

Condition (12) accounts for the above relation between the shear stress along the high-angle border and the pressure affecting it.

The plate and the inclusion are made of the same material that is an isotropic elastic body with elastic modulus $E$ and Poisson's ratio $\nu$.

We numerically analyzed planar deformation of the system under simple shear (loading 1) and under lengthening along the side $B$ (loading 2). The following parameters were used: $D = 1$, $H = 10$, $B = 100$, $\nu = 0.3$, $\tau_0 = 5 \cdot 10^{-4} E$, $\mu = 10^{-3}$. Linear sizes satisfy conditions $H/B \ll 1$ and $D/H \ll 1$ that exclude edge effects. Values that were chosen for $\nu$ and $\tau_0/E$ are typical for metals, $\mu$ was estimated based on the relationship between the shear stress and pressure for metals [2, 26].

In order to establish specifics of each mode of deformation, during the loading the same maximum shear $\gamma$ was achieved. Under the simple shear scheme necessary shift $\Delta$ of the upper plate relative to the lower one was determined by:

$$\Delta = \gamma H \qquad (13)$$

Under the second loading scheme the necessary lengthening of the plate $\Delta B$ was given by the relation between maximum lengthening and maximum shear [27]. As a result the following expression was derived:

$$\Delta B = \gamma B(1 - \nu) \qquad (14)$$

Simulations were made for the shear range $0 \leq \gamma \leq 0.005$ employing ANSYS package. The finite element PLANE183 is used. This element is defined by 8-nodes having two degrees of freedom at each node: translations in the nodal x and y directions.

In order to correctly simulate the problem, a contact analysis is utilized. For this purpose contact elements CONTA172 are placed along the matrix surface and target elements TARGE169 are used along the surface of inclusion. For surface-to-surface contact elements the Lagrange multiplier method on contact normal and the penalty method on tangential contact stiffness are used. This method enforces zero penetration and allows a small amount of slip for the sticking contact. The amount of slip in sticking contact depends on the tangential stiffness.

Figure 8 illustrates typical vector plot of displacements for deformation under two schemes.

Numerical modeling showed that loading of the system under two schemes have similarities as well as differences in the behavior of inclusion. Similarity is that beginning with deformation value $\gamma^* \sim 10^{-3}$ along certain areas of the border slipping occurs. When deformation increases the number of such areas and the magnitude of the slip also increase. The difference between the loading schemes is that under the lengthening scheme slipping is rare and is of accidental nature, while under simple shear it is coordinated and lowers the shear deformation of the inclusion (Figure 9).

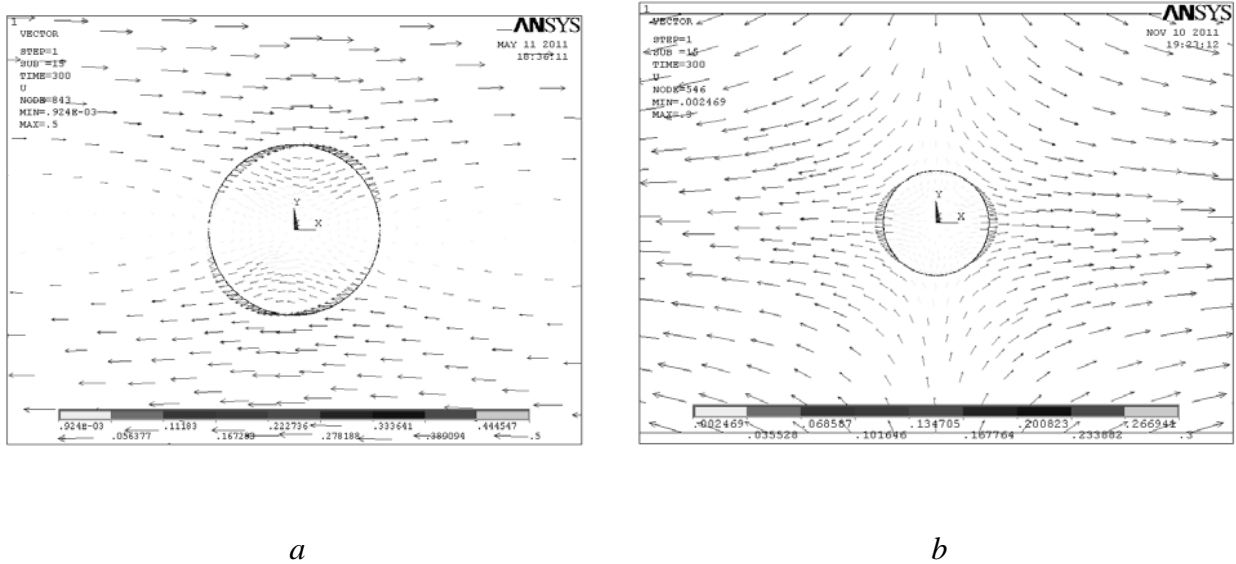

*a*                                         *b*

Figure 8. Typical vector plots of displacement for deformations under simple shear (a) and flat lengthening schemes (b)

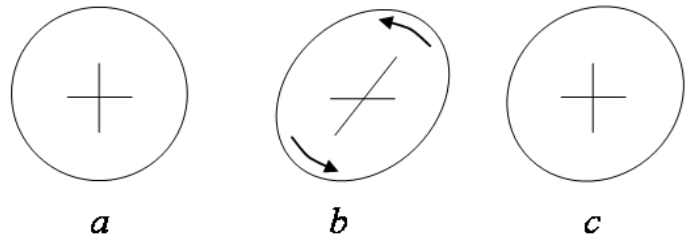

*a*            *b*            *c*

Figure 9. The scheme describes directional nature of slipping along the border between inclusion and the plate under simple shear: inclusion prior to deformation (a), after simple shear without slipping (b), after slipping along the border (c). Deformations are significantly exaggerated for the sake of clearness.

Inclusion rotates and deforms in such a way that elastic energy of the entire system decreases.

This result is easy to understand qualitatively. Shear deformation of inclusion is related with elastic energy given by $W \sim V E \gamma^2$, where $V = \dfrac{\pi D^2}{4}$ is the volume of the inclusion (thickness is assumed to be unit). Correspondingly, an increase in elastic energy is equal to $\Delta W \sim L^2 E \gamma \Delta \gamma$. Rotation occurs when the accompanying decrease in elastic energy exceeds the work of the friction force $\Delta A$ along the border of the inclusion, which can be assessed as $\Delta A \sim L^2 \tau_0 \Delta \gamma$. So the condition of the rotation is $\gamma^* \sim \dfrac{\tau_0}{E}$, which by the order of value complies with $\gamma^*$ assessed empirically.

## 6. Discussion

In previous sections we tried to justify a hypothesis that perfect plasticity under low homological temperatures is a critical phenomenon, which is incident to metals only under simple shear. Starting from a certain stage of deformation in the grain boundaries lattice percolation clusters emerges – they provide the simple shear through the representative volume. Percolation is a collective effect, so it is the cooperative mechanism that is responsible for perfect plasticity. As it was shown in sections 4 and 5, it is a multi-scale rotation of the material blocks in percolation clusters cells.

Another, local, variant of perfect plasticity mechanism is suggested and grounded in [5-7], where this phenomenon is connected with grains boundaries migration. Under this approach deformation mode does not influence the evolution of metals microstructure and perfect plasticity should emerge under any loading scheme. As a direct argument in favor of their hypothesis the authors of [5-7] give the results of following experiment carried out by means of High Pressure Torsion (HPT) method.

In a thin nickel disc by high pressure torsion a stationary structure was created that induced perfect plasticity. Afterwards, on the parallel to the axle crosscut of the specimen a squared grid was dashed by a high-energy ionic beam. Then the disc was twisted once more so that the shear deformation in the crosscut was ~1. According to authors, nearly homogeneous grid deformation testified against the grain boundaries slipping and fragmentation at the stationary stage of deformation stopped due to grains boundaries migration.

From our point of view the described experiment does not allow even to make a conclusive implication whether the mechanisms of perfect plasticity under low homological temperatures are local. There are at least two factors that could in this case result in that the grid did not spread out under deformation as it should have been according to property (b) in section 4.

The first one is that high-energy ionic beam could have changed the local structure of the boundaries on the grid lines, so that slipping on them became impossible. Hard influence of such beams on the material is documented in numerous works (e.g. in [28]).

The second factor is that in the thin butt end of the disc with the crosscut, containing the grid, deformation mode changed. Absence of pressure on the crosscut surface resulted in flow of the material in radial direction. This, in turn, reduced the thickness of the disc in the crosscut area and, consequently, the pressure of anvils on it. Eventually, the thin layer of the disc containing the grid turned into a kind of sticker on the disc. Its deformation replicated the disc deformation, but the mechanism of deformation did not correspond to the one for the stationary stage of simple shear.

It should be noted that there has been a body of convincing experimental evidence about abnormally fast mass transfer in metals under simple shear [12-15] as it should have been according to suggested cooperative mechanism (property (b) in section 4).

An additional point to emphasize is that activation energy for grained boundaries migration is considerably above the activation energy for grained boundaries sliding (e.g. in [20, 21]). Grained boundaries migration at low homological temperatures may be a consequence of grain boundary sliding due to boundaries steps.

Arguments in favor of suggested perfect plasticity mechanism are adduced by some well known experimental results.

At the stationary HPT stage the average fragments size, the size distribution of the fragments and part of high-angle grain boundaries do not change as deformation increases.

Values of theses parameters correspond to the ones at the terminating of fragmentation [4, 29]. This clearly demonstrates the property (c) of suggested perfect plasticity mechanism.

HPT experiments with Fe specimens under von Mises strain $e = 300$ demonstrated that material fragments on the perpendicular to the radius crosscut were slightly extended and tilted at an angle to an anvil axis smaller than evident from geometry one (70-80 degrees instead of 89.89). The real tilt of the fragments was appropriate to deformation $e = 1.6 \div 3.3$ [5]. Property (c) in section 4 offers a satisfactory explanation of the observed fact.

It has been apparent in recent reports on severe plastic deformation (SPD) under simple shear scheme that scaling effect exists. It lies in the fact that the average grain size of submicrocrystalline structure undergoes a rise as the size of specimen increases (at the same another conditions). Scaling effect has come to light in the examination of Equal Channel Angular Pressing (ECAP) [30]. It is also found in Twist Extrusion (TE) experiments [31]. Within the context of suggested theory the property (a) in section 4 is associated with SPD scaling effect. From relation (9) it is apparent that to enhance the part of high-angle grain boundaries at steady-state phase of simple shear the size of specimen should be decreased. It is not proof that perfect plasticity has percolation nature so SPD scaling effect requires further examination. But if the only distinctive structure scale is the average grain size of order $100 nm$ then it is difficult to explain that the thickness of the shear layer of four orders of magnitude higher has an impact on it.

Finally, we note that the proposed mechanism of perfect plasticity is implemented by a relatively small elastic-plastic deformation of the volume of material bounded by the cells of a percolation cluster. Moreover, these strains are cyclical in nature. At this stage fragments get free of dislocations (they get beyond the boundaries), microvoids that formed at the early stages of deformation are healed. As a result the ductility of the metal enhances [15, 32].

## 7. Conclusion

This article attempts to explain the nature of perfect metal plasticity at low homological temperatures. There are reasons to assume that it is realized only under simple shear and performed by cooperative mechanism due to percolation shift along grain boundaries. Prior to some percolation threshold metal deformation under simple shear is realized by the same mechanism as under strengthening. In this case both of these deformation modes make an equivalent impact on metal hardening rule and grain refinement.